\documentclass[aps,superscriptaddress,twocolumn,nofootinbib]{revtex4}
\input epsf

\usepackage{natbib}
\usepackage{graphicx}
\usepackage{dcolumn}
\begin{document}
\title{The Predictive Power of Zero Intelligence in Financial Markets}
\author{J. Doyne Farmer} 
\email{jdf@santafe.edu} 
\affiliation{Santa Fe Institute, 1399 Hyde Park Road, Santa Fe, NM 87501, USA}
\author{Paolo Patelli}
\affiliation{Santa Fe Institute, 1399 Hyde Park Road, Santa Fe, NM 87501, USA}
\affiliation{Sant'Anna School of Advanced Studies, Piazza Martiri della Libert\`a 33, Pisa, Italy}
\author{Ilija I. Zovko} 
\affiliation{Santa Fe Institute, 1399 Hyde Park Road, Santa Fe, NM 87501, USA}
\affiliation{CENDEF, University of Amsterdam, Roetersstraat 11, Amsterdam, The Netherlands}

\date{Original version Sept. 9, 2003; this version Feb. 9, 2004}

\begin{abstract}

Standard models in economics stress the role of intelligent agents who
maximize utility.  However, there may be situations where, for some
purposes, constraints imposed by market institutions dominate
intelligent agent behavior.  We use data from the London Stock
Exchange to test a simple model in which zero intelligence agents
place orders to trade at random. The model treats the statistical
mechanics of order placement, price formation, and the accumulation of
revealed supply and demand within the context of the continuous double
auction, and yields simple laws relating order arrival rates to
statistical properties of the market.  We test the validity of these
laws in explaining the cross-sectional variation for eleven
stocks. The model explains 96\% of the variance of the bid-ask spread,
and 76\% of the variance of the price diffusion rate, with only one
free parameter.  We also study the market impact function, describing
the response of quoted prices to the arrival of new orders.  The
non-dimensional coordinates dictated by the model approximately
collapse data from different stocks onto a single curve.  This work is
important from a practical point of view because it demonstrates the
existence of simple laws relating prices to order flows, and in a
broader context, because it suggests that there are circumstances
where institutions are more important than strategic considerations.
\end{abstract}
\maketitle 

\tableofcontents

\section{Introduction}

This work has goals at two levels.  At the immediate level, its
goal is to investigate the possibility of simple laws relating the
flow of trading orders into a market to statistical properties of
prices.  The laws that we propose and investigate are not temporal
predictions, but rather relations restricting the possible
values that the underlying variables can take at any given point in
time.  The ideal gas law provides a simple physical analogy that
illustrates both the limited scope and the potential utility of such
laws.  In our case, the goal is to relate properties of the order
flow, such as market order placement rate, limit order placement rate,
and cancellation rate, to properties of the market such as the gap
between the best prices for buying and selling, or the variability of
prices.  In addition, we present some results that are related to the
nature of supply and demand functions.

At a broader level, this work is interesting because of the nature of
the model we test, which makes the simple assumption that agents place
orders to buy or sell at random \cite{Daniels03,Smith03}.  This is in
constrast to standard models in economics, which typically devote
considerable effort to modeling the strategic behavior and
expectations of agents.  No one would dispute that this is important.
However, there may be some circumstances where other factors may be
more important.  For example, Becker \cite{Becker62} showed that a
budget constraint is sufficient to guarantee the proper slope of
supply and demand curves, and Gode and Sunder \cite{Gode93}
demonstrated that if one replaces the students in a standard classroom
economics experiment by zero-intelligence agents, the
zero-intelligence agents perform surprisingly well. The model we test
here builds on earlier work in financial economics
\cite{Mendelson82,Cohen85,Domowitz94,Bollerslev97} and physics
\cite{Bak97,Eliezer98,Maslov00,Slanina01,Challet01}. (See also
interesting subseqent work \cite{Bouchaud02,Bouchaud03}).  We show
here that in some circumstances the zero-intelligence approach can
make surprisingly good quantitative predictions.

\subsection{Continuous double auction}

\begin{figure}[bt]
   \begin{center}
    \includegraphics[scale=0.4]{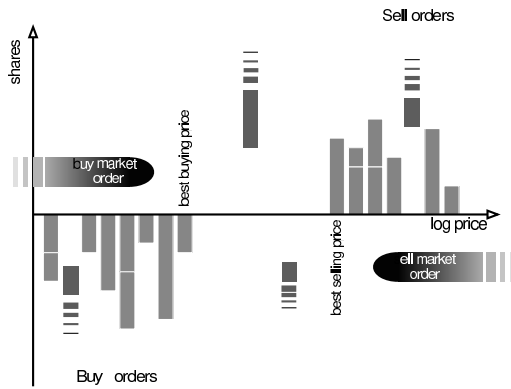}
    \caption{\small A random process model of the continuous double
    auction. Stored limit orders are shown stacked along the price
    axis, with sell orders (supply) stacked above the axis at higher
    prices and buy orders (demand) stacked below the axis at lower
    prices.  New sell limit orders are visualized as randomly falling
    down, and new buy orders as randomly ``falling up''.  New sell
    orders can be placed anywhere above the best buying price, and
    new buy orders anywhere below the best selling price.  Limit
    orders can be removed spontaneously (e.g. because the agent
    changes her mind or the order expires) or they can be removed by
    market orders of the opposite type.  This can result in changes
    in the best prices, which in turn alters the boundaries of the
    order placement process. It is this feedback between order
    placement and price formation that makes this model interesting,
    and its predictions non-trivial.}
       \label{orderSchematic} 
\end{center}
\end{figure}

The model of Daniels et al. \cite{Daniels03} assumes a continuous
double auction, which is the most widely used method of price
formation in modern financial markets \cite{Smith03}.  There are two
fundamental kinds of trading orders: Impatient traders submit
\emph{market orders}, which are requests to buy or sell a desired
number of shares immediately at the best available price.  More
patient traders submit \emph{limit orders}, which include the worst
allowable price for the transaction.  Limit orders may fail to result
in an immediate transaction, in which case they are stored in a queue
called the \emph{limit order book}, illustrated in
Fig.~\ref{orderSchematic}.  As each buy order arrives it is transacted
against accumulated sell limit orders that have a lower selling price,
in priority of price and arrival time.  Similarly for sell orders.
The lowest selling price offered in the book at any point in time is
called the \emph{best ask}, $a(t)$, and the highest buying price the
\emph{best bid}, $b(t)$.

\subsection{Review of model}

The model that we test here \cite{Daniels03,Smith03} assumes that two
types of zero intelligence agents place and cancel orders randomly, as
shown in Fig.~\ref{orderSchematic}.  Impatient agents place market
orders of size $\sigma$, which arrive at a rate $\mu$ \emph{shares per
  time}.  Patient agents place limit orders of the same size $\sigma$,
which arrive with a constant rate density $\alpha$ \emph{shares per
  price per time}.  These agents may be thought of as liquidity
demanders and suppliers.  Queued limit orders are canceled at a
constant rate $\delta$, with dimensions of \emph{1/time}. Prices
change in discrete increments called \emph{ticks}, of size $dp$.  To
keep the model as simple as possible, there are equal rates for buying
and selling, and order placement and cancellation are Poisson
processes.  All of these processes are independent except for coupling
through their boundary conditions: Buy limit orders arrive with a
constant density $\alpha$ over the semi-infinite interval $-\infty < p
< a(t)$, where $p$ is the logarithm of the price, and sell limit
orders arrive with constant density $\alpha$ on the semi-infinite
interval $b(t) < p < \infty$. As a result of the random order arrival
processes, $a(t)$ and $b(t)$ each make random walks, but because of
coupling of the buying and selling processes the bid-ask \emph{spread}
$s(t) \equiv a(t) - b(t)$ is a stationary random variable.

As new orders arrive they may alter the best prices $a(t)$ and $b(t)$,
which in turn changes the boundary conditions for subsequent limit
order placement. For example, the arrival of a buy limit order inside
the spread will alter the best bid $b(t)$, which immediately alters
the boundary condition for sell limit order placement.  It is this
feedback between order placement and price diffusion that makes this
model interesting, and despite its apparent simplicity, quite
difficult to understand analytically.  This model has been analyzed
using both simulation and two different mean field
theories~\cite{Smith03}.

One of the virtues of this model is that it gives simple scaling laws
relating the parameters of the model to fundamental properties such as
the average bid-ask spread, and the price diffusion rate.  The mean
value of the spread predicted based on a mean field theory analysis of
the model in the limit $dp \rightarrow 0$ is
\begin{equation}
\label{spreadScaling}
\hat{s} = (\mu/\alpha) f(\sigma \delta/\mu).
\end{equation}
The nondimensional ratio $\epsilon = \sigma \delta/\mu$ is the ratio
of removal by cancellation to removal by market orders, and plays an
important role in determining the properties of the model.
$f(\epsilon)$ is a relatively slowly varying, monotonically increasing
non-dimensional function that can be approximated as $f(\epsilon) =
0.28 + 1.86 \epsilon^{3/4}$.

Another prediction of the model is of the price diffusion rate, which
drives the volatility of prices and is the primary determinant of
financial risk.  If we assume that prices make a random walk, then the
diffusion rate measures the size and frequency of its increments. The
variance $V$ of an uncorrellated normal random walk after time $t$
grows as $V(t) = Dt$, where $D$ is the diffusion rate. We choose to
measure the price diffusion rate rather than the volatility because it
is a stationary quantity that provides a more fundamental description
of the volatility process. This is the main free parameter in the
Bachelier model \cite{Bachelier00}, and while its value is essential
for risk estimation and derivative pricing there is very little
understanding of what determines it.  Numerical experiments indicate
that the short term price diffusion rate predicted by the model is
\begin{equation}
\label{volScaling}
        \hat{D} = k \mu^{5/2} \delta^{1/2} \sigma^{-1/2} \alpha^{-2},
\end{equation}
where $k$ is a constant. 

The model was constructed to be simple enough to be analytically
tractable, and so makes many strong assumptions.  For example, it
assumes that the rates for buying and selling are equal, the sizes of
limit orders and market orders are the same, that limit order
deposition is uniform on semi-infinite intervals, and that rates
of order submission are unaffected by changes in price.  Many of these
assumptions are economically unreasonable in the presence of
intelligent agents, but the reader should bear in mind that the only
market participants in the model are zero-intelligence ``noise''
traders, who can be thought of as random liquidity suppliers and
demanders\footnote{A ``liquidity demander'' is someone who needs to
  make a transaction quickly.  In the sense used here, a noise trader
  is someone who wants to make transactions for reasons unrelated to
  this particular market, and so is insensitive to price.}.  While
intelligent agents are clearly essential for many purposes, such as
determining the levels of prices, what we suggest here is that for
other purposes their presence is not essential.  We would like to
emphasize that the construction of the model and all the predictions
derived from it were made prior to looking at the data.

\section{Testing the scaling laws}

\subsection{Data}

We test this model with data from the electronic open limit order book
of the London Stock Exchange (SETS), which includes about half of the
total volume on the exchange.  We used data from eleven stocks for
August 1st 1998 - April 30th 2000, which includes 434 trading days and
a total of roughly six million events.  For all these stocks the
number of total events exceeds 300,000 and was never less than 80 on
any given day (where an event corresponds to an order placement or
cancellation).  Orders placed during the opening auction are removed
to accomodate the fact that the model only applies for the continuous
auction.  See the Supplementary Material Section~\ref{datasetDesc} for
more details.

\subsection{Testing procedure}

>From the point of view of the model, the order flow rates $\mu$,
$\alpha$, and $\delta$, and the mean order size $\sigma$ are all free
parameters.  In analyzing the model we find scaling relations
connecting these parameters to the average spread and the price
diffusion rate, as given in Equations~\ref{spreadScaling} and
\ref{volScaling}.  We test the model by testing the validity of these
relations, taking advantage of the fact that different stocks have
different average values of these parameters.  For each stock we
measure the average market order arrival rate $\mu$, limit order
rate density $\alpha$, cancellation rate $\delta$, and order size $\sigma$,
where the averages are taken across the full time period.  We then
measure the average spread and volatility and compare them to the
predictions of the model.

A problem occurs in measuring $\alpha$ and $\delta$ due to the
simplifying assumption of a uniform distribution of prices for order
flow and cancellation.  In the real data order placement and
cancellation are concentrated near the best prices
\cite{Bouchaud02,Zovko02}.  We cope with this by making the assumption
that order placement is uniform inside a price window around the best
prices, and zero outside this window.  We choose the price window to
correspond to roughly $60\%$ of limit orders at the best prices, and
compute $\alpha$ by dividing the number of shares of limit orders
placed inside the price window by the size of the price window.  We
do this for each day and compute the average value of $\alpha$ for
each stock.  We compute $\delta$ as the inverse of the average
cancellation time for orders cancelled inside the same price window.
See the Supplementary Material Section~\ref{param_measurement} for
details.

The scaling laws that we describe here do not make temporal
predictions, but rather are restrictions of state variables.  The
ideal gas law, $PV = RT$, provides a good analogy.  It predicts that
pressure $P$, volume $V$, and temperature $T$ are constrained -- any
two of them determines the third.  Similarly, here we are testing two
relations relating properties of orders to properties of prices.  We
are not attempting to predict the temporal behavior of the order
flows, only trying to see whether the restrictions between order flows
and prices are valid. 

We would like to emphasize that in testing the model we are not
treating the order flow rates and order size as free parameters in the
regressions.  Instead, we are testing the predictions of the model
based on order flow rates against the measured values in the same
period.  The only free parameters are in the specification of the
price interval as described above (which was done more or less
arbitrarily).

\subsection{Spread}

To test Equation~\ref{spreadScaling}, we measure the average
spread $\bar{s}$ across the full time period for each stock, and
compare to the predicted average spread $\hat{s}$ based on order
flows.  Spread is measured as the daily average of $\log b(t) -
\log a(t)$.  The spread is measured after each event, with each
event given equal weight.  The opening auction is excluded.

To test our hypothesis that the predicted and actual values coincide,
we perform a regression of the form $\log \bar{s} = A \log \hat{s} +
B$.  We used logarithms because the spread is positive and the log of
the spread is approximately normally distributed.  We use the free
parameters $A$ and $B$ for hypothesis testing.  Based on the model we
predict that the comparison should yield a straight line with $A=1$
and $B=0$, but because of the degree of freedom in choosing the price
interval as described above, the value of $B$ is somewhat arbitrary.

\begin{figure}[!tbh]
   \begin{center} 
   \includegraphics[scale=0.4]{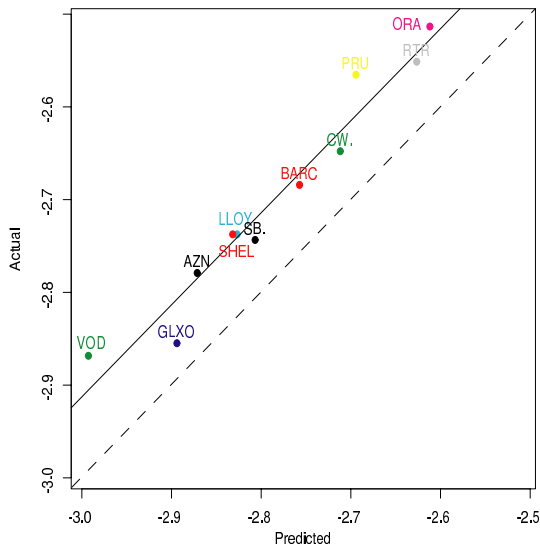}
 \vskip -2cm 
    \caption{\small Regressions of predicted values based on order
      flow parameters vs. actual values for the log spread.  The dots
      show the average predicted and actual value for each stock
      averaged over the full 21 month time period.  The solid line is
      a regression; the dashed line is the diagonal, representing the
      model's prediction without any adjustment}
  \label{spreadRegression}
  \end{center}
\end{figure}

The least squares regression, shown together with the data comparing
the predictions to the actual values in Fig.~\ref{spreadRegression},
gives $A = 0.99 \pm 0.10$ and $B = 0.06 \pm 0.29$.  We thus strongly
reject the null hypothesis that $A = 0$, indicating that the
predictions are far better than random.  More importantly, we are
unable to reject the null hypothesis that $A = 1$. In fact, we are
also unable to reject $B = 0$, but this is probably largely a matter
of luck in our choice of the price interval. The regression has $R^2 =
0.96$, so the model explains most of the variance.  Note that because
of long-memory effects and cross-correlations between stocks the
errors in the regression are larger than they would be for IID data
(see the discussion in the Supplemenary Material
Section~\ref{regressionErrors}).

\subsection{Price diffusion rate}

\begin{figure}[htb]
   \begin{center} 
   \includegraphics[scale=0.4]{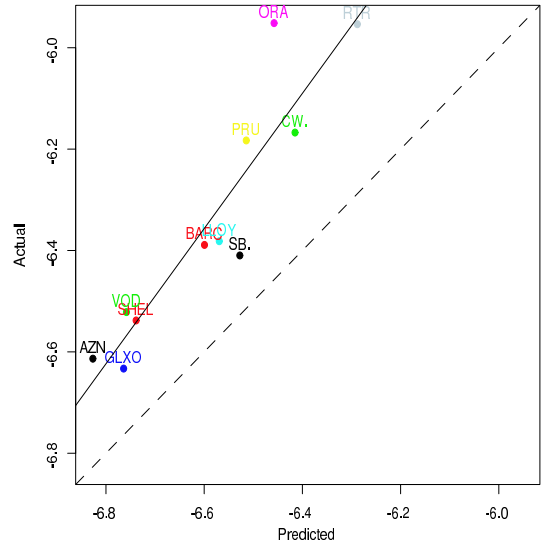}
 \vskip -2cm 
    \caption{\small Regressions of predicted values based on order
      flow parameters vs. actual values for the logarithm of the price
      diffusion rate.  The dots show the average predicted and actual
      value for each stock averaged over the full 21 month time
      period.  The solid line is a regression; the dashed line is the
      diagonal, representing the model's prediction without any
      adjustment of slope or intercept.}
  \label{diffusionRegression}
  \end{center}
\end{figure}

As for the spread, we compare the predicted price diffusion rate based
on order flows to the actual price diffusion rate $\bar{D}_i$ for each
stock averaged over the 21 month period, and regress the logarithm of
the predicted vs. actual values, as shown in
Fig.~\ref{diffusionRegression}.

The regression gives $A = 1.33 \pm 0.25$ and $B = 2.43 \pm 1.75$.
Thus, we again strongly reject the null hypothesis that $A = 0$.  We
are still unable to reject the null hypothesis that $A = 1$ with
$95\%$ confidence, though there is some suggestion that the scaling of
the model and the actual values are not quite the same.  (This could
happen if, for example, the scaling exponent predicted by the model of
one or more of the order flow rates is wrong; however this suggests
that it is at least quite close).  Although the results are not as
good as for the spread, $R^2 = 0.76$, so the model still explains most
of the variance.

\section{Average market impact}

Market impact is practically important because it is the dominant
source of transaction costs for large traders, and conceptually
important because it provides a convenient probe of the revealed
supply and demand functions in the limit order book.  When a market
order of size $\omega$ arrives, if sufficiently large, it will remove
all the stored limit orders at the best bid or ask, causing a change
in the midpoint price $m(t) \equiv (a(t) + b(t))/2$. The average
market impact function $\phi$ is the average logarithmic midpoint
price shift $\Delta p$ conditioned on order size, $\phi(\omega) =
E[\Delta p | \omega]$.

A long-standing mystery about market impact is that it is highly
concave
\cite{Hausman92,Farmer96,Torre97,Kempf98,Plerou01,Bouchaud02,Lillo03,Gabaix03}. This
is unexpected since simple arguments would suggest that because of the
multiplicative nature of returns, market impact should grow at least
linearly \cite{Smith03}.  The model we are testing predicts a concave
average market impact function, with the concavity becoming more
pronounced for small values of $\epsilon = \sigma \delta/\mu$.
However, these predictions are not in good detailed agreement with the
data, in that the model predicts a larger variation with $\epsilon$
than what is actually observed.  However, the model is still useful
for understanding market impact, as described below.

\subsection{Collapse in non-dimensional coordinates}

A surprising regularity of the average market impact function is
uncovered by simply plotting the data in non-dimensional coordinates,
as shown in Fig.~\ref{marketImpactPic}.  See the Supplementary
Material Section~\ref{modelBackground} for a discussion of how the
nondimensional coordinates are derived from the model.  Each market
order $\omega_i$ causes a possible change $\Delta p_i$ in the midquote
price.  If we bin together events with similar $\omega$ and plot the
mean order size as a function of the mean price impact $\Delta p$, we
typically see highly variable behavior for different stocks, as shown
in Fig.~\ref{marketImpactPic}(b).  We have also explored other ways of
renormalizing the order size, such as taking the ratio of each order's
size to the daily or full-sample mean, but they give similar behavior,
as shown in the Supplementary Material Section~\ref{marketImpact}.

Plotting the data in non-dimensional units tells a simpler story. This
involves normalizing the price shift and order size by appropriate
dimensional scale factors based on the daily order flow rates.  In
particular, $\Delta p \rightarrow \Delta p \alpha_t /\mu_t$ and
$\omega \rightarrow \omega \delta_t/\mu_t$, where $\alpha_t$, $\mu_t$,
and $\delta_t$ are the average order flow rates for day $t$. The data
collapses onto roughly a single curve, as shown in
Fig.~\ref{marketImpactPic}(a).  The variations from stock to stock are
quite small; on average the corresponding bins for each stock deviate
from each other by about $8\%$, roughly the size of the statistical
sampling error. We have made an extensive analysis, but due to
problems caused by the long-memory property of these time series and
cross correlations between stocks, it remains unclear whether these
differences are statistically significant.  In contrast, using
standard coordinates the differences are highly statistically
significant. This collapse illustrates that the non-dimensional
coordinates dictated by the model provide substantial explanatory
power: We can understand how the average market impact varies from
stock to stock by a simple transformation of coordinates. Plotting in
double logarithmic scale shows that the curve of the collapse is
roughly a power law of the form $\omega^{0.25}$~(see Supplementary
Material, Section~\ref{marketImpact}). This provides a more
fundamental explanation for the empirically constructed collapse of
average market impact for the New York Stock Exchange found
earlier~\cite{Lillo03}.

\begin{figure}[htb]
   \begin{center}
   \includegraphics[scale=0.4]{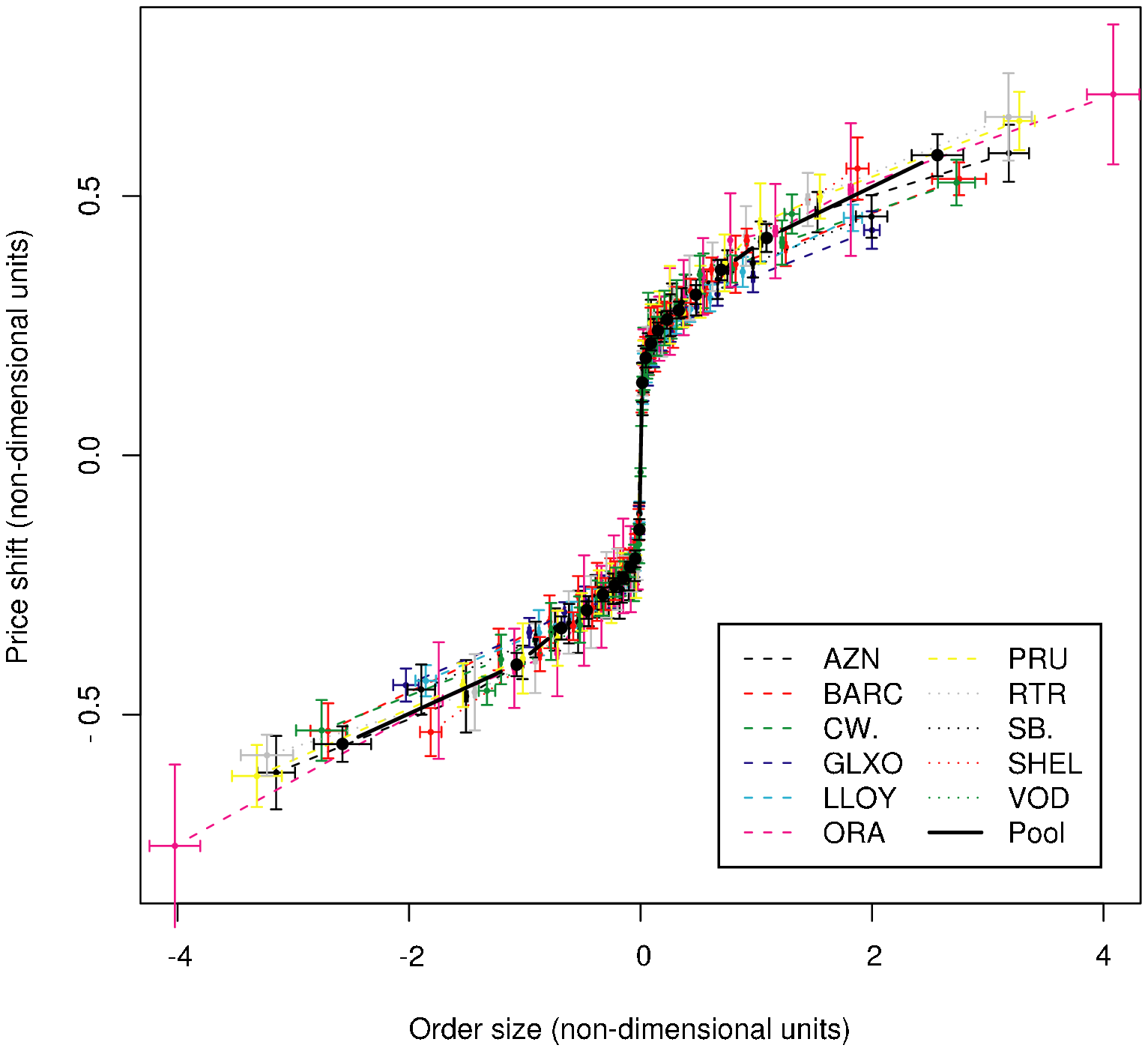} 
\vskip -2cm
   \includegraphics[scale=0.4]{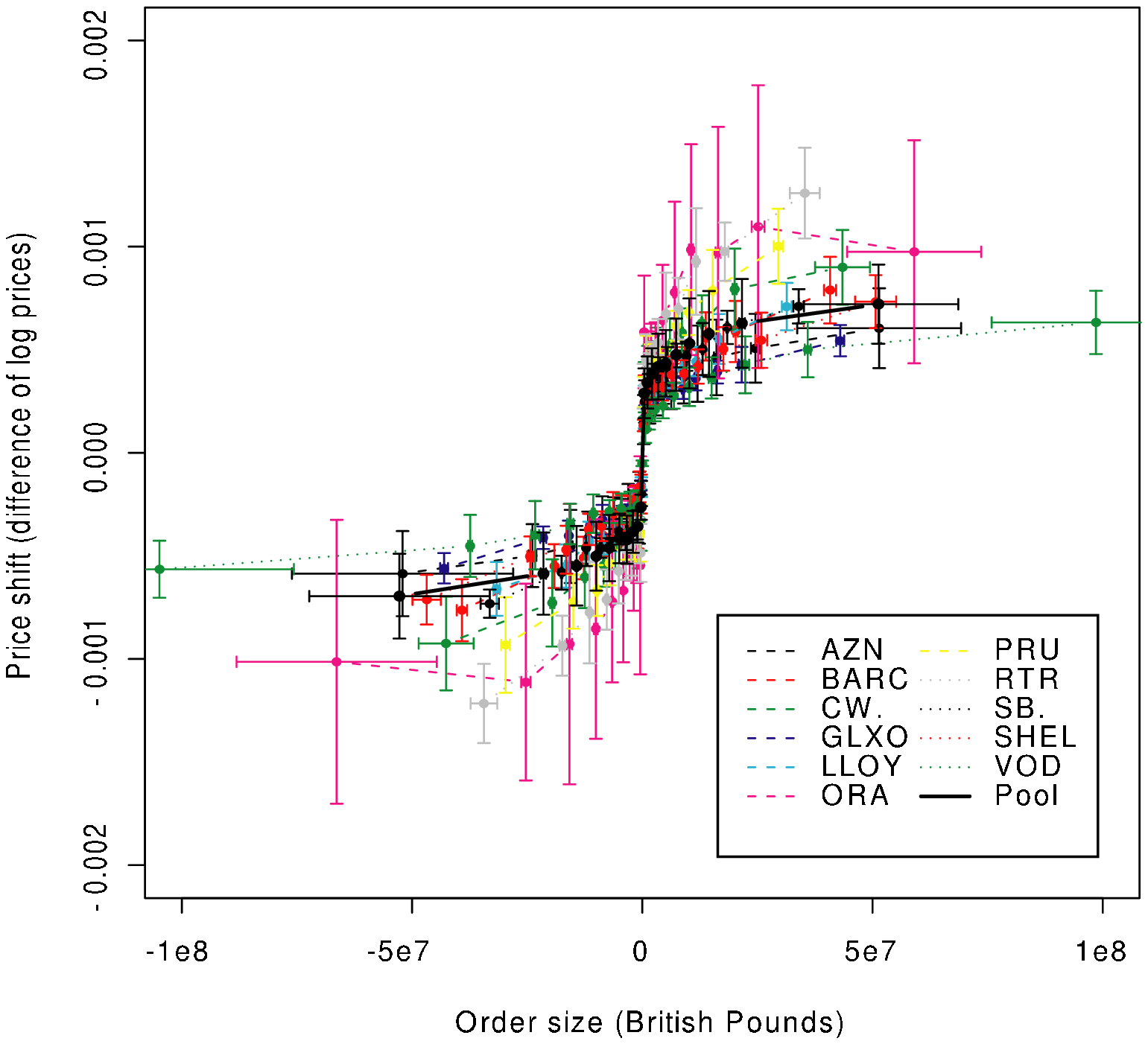} 
\vskip -3cm
   \end{center} 
\caption{\small The average market impact as a function of the mean
  order size.  In (a) the price differences and order sizes for each
  transaction are normalized by the non-dimensional coordinates
  dictated by the model, computed on a daily basis.  Most of the
  stocks collapse extremely well onto a single curve; there are a few
  that deviate, but the deviations are sufficiently small that given
  the long-memory nature of the data and the cross-correlations
  between stocks, it is difficult to determine whether these
  deviations are statistically significant.  This means that we
  understand the behavior of the market impact as it varies from stock
  to stock by a simple transformation of coordinates.  In (b), for
  comparison we plot the order size in units of British pounds against
  the average logarithmic price shift.}
  \label{marketImpactPic}
\end{figure}

\section{Conclusions}

The model we have presented here does a good job of predicting the
average spread, and a decent job of predicting the price diffusion
rate.  Also, by simply plotting the data in non-dimensional
coordinates we get a better understanding of the regularities of
market impact.  These results are remarkable because the underlying model
completely drops agent rationality, instead focusing all its
attention on the problem of understanding the constraints imposed by
the continuous double auction.

The approach taken here can be viewed as a divide and conquer
strategy.  Rather than attempting to explain the properties of the
market from fundamental assumptions about utility maximization by
individual agents, we divide the problem into two parts. The first and
much easier problem, addressed here, is that of understanding the
characteristics of the market given the order flows.  The second (and
harder) problem, which remains to be investigated, is that of
explaining why order flow varies as it does.  Explaining order flow
involves behavioral and/or strategic issues that are likely to be much
more difficult to understand.

The model that we test succeeds in part because it takes explicit
advantage of information that is available in a continuous double
auction, that is not available in a standard Walrasian auction.  By
measuring the rate of market order placement vs. limit order
placement, and the rate of order cancellation, we are able to measure
how patient or impatient traders are.  A higher ratio of market orders
to limit orders, or a higher rate of cancellation implies a less
patient, and therefore more volatile market, with larger spreads.  The
model makes this quantitative.  The agreement with the model indicates
that the degree of patience is an important determinant of market
behavior.  This is potentially compatible with either a
rationality-based explanation in terms of information arrival, or a
behavioral-based explanation driven by emotional response, but in
either case it suggests that patience is a key factor.

This is part of a broader research program that might be characterized
as the ``low-intelligence'' approach to economics: We begin with
zero-intelligence agents to get a good benchmark of the effect of
market institutions, and once this benchmark is well-understood, add a
little intelligence, moving toward market efficiency.  We thus start
from zero rationality and work our way up, in contrast to the
canonical approach of starting from perfect rationality and working
down.  Follow-up research will examine the effects of adding
bounded rationality.  See Ref.~\cite{Farmer03c}.

These results have several practical implications.  For market
practitioners, understanding the spread and the market impact function
is very useful for estimating transaction costs and for developing
algorithms that minimize their effect.  For regulators they suggest
that it may be possible to make prices less volatile and lower
transaction costs by creating incentives for limit orders and
disincentives for market orders.  These scaling laws might be
used to detect anomalies, e.g. a higher than expected spread might be
due to improper market maker behavior.

The model we test here was constructed before looking at the data
\cite{Daniels03,Smith03}, and was designed to be as simple as possible
for analytic analysis.  A more realistic (but necessarily more
complicated) model would more closely mimic the properties of real
order flows, which are price dependent and strongly correlated both in
time and across price levels, or might incorporate elements of the
strategic interactions of agents.  An improved model would hopefully
be able to capture more features of the data than those we have
studied here.  We know there are ways in which the current model is
inappropriate, e.g., predicts unrealistically strong negative
autocorrelations in prices, allowing arbitrage opportunities that do
not exist in the real market.  Nonetheless, as we have shown above,
this extremely simple model does a good job of explaining some
important properties of markets, such as transaction costs, price
diffusion and market impact.  It does this by focusing on the way
order placement and price formation interact to alter the accumulation
of stored supply and demand.  For the phenomena studied here this
appears to be the dominant effect.  We do not mean to claim that
market participants are unintelligent: Indeed, one of the virtues of
this model is that it provides a benchmark to separate properties that
are driven by the statistical mechanics of the market institution from
those that are driven by conditional strategic behavior. It is
surprising that such a simple model can explain anything at all about
a system as complex as a market.

\vskip 1cm
We would like to thank Credit Suisse First Boston, the James S. McDonnell
Foundation, the McKinsey Corporation, Bob Maxfield, and Bill Miller
for their support of this project.  We would also like to thank Sam
Bowles, Supriya Krishnamurthy, Fabrizio Lillo, and Eric Smith for
useful discussions, and Mark Bieda, David Krakauer, Harold Morowitz
and Elizabeth Wood for comments on the manuscript.

\clearpage

\begin{center}

\end{center}

\appendix

\section{Supplementary Material}

\subsection{Additional background information on the model}
\label{modelBackground}

One of the virtues of this model is that we can make approximate
predictions of several of its properties with almost no work using
dimensional analysis.  This also greatly simplifies the analysis and
understanding of the model, and is particularly useful for
understanding market impact.  

There are three fundamental dimensional quantities describing
everything in this model: \emph{shares}, \emph{price}, and
\emph{time}.  There are five parameters defined in the model.  When
the dimensional constraints between the parameters are taken into
account, this leaves only two independent degrees of freedom.  It
turns out that the order flow rates $\mu$, $\alpha$, and $\delta$ are
more important than the discreteness parameters $\sigma$ and $dp$, in
the sense that the properties of the model are much more sensitive to
variations in the order flow rates than they are to variations in
$\sigma$ or $dp$.  It therefore natural to construct non-dimensional
units based on the order flow parameters alone.  There are unique
combinations of the three order flow rates with units of
\emph{shares}, \emph{price}, and \emph{time}.  This gives
characteristic scales for price, shares, and time, that are unique up
to a constant.  In particular, the characteristic number of shares
$N_c = \mu / \delta$, the characteristic price interval $p_c = \mu /
\alpha$, and the characteristic timescale $t_c = 1/\delta$.

These characteristic scales can be used to define non-dimensional
coordinates based on the order flow rates.  These are $\hat{p} =
p/p_c$ for price, $\hat{N} = N/N_c$ for shares, and $\hat{t} = t/t_c$
for time.  The use of non-dimensional coordinates has the great
advantage that it reduces the number of degrees of freedom from five
to two, and many quantities are much more well-behaved
and easily understood when plotted in non-dimensional coordinates than
they are otherwise.

The remaining two degrees of freedom are naturally discussed in terms
of non-dimensional versions of the discreteness parameters.  A
non-dimensional scale parameter based on order size is constructed by
dividing the typical order size $\sigma$ (with dimensions of
\emph{shares}) by the characteristic number of shares $N_c$.  This
gives the non-dimensional parameter $\epsilon \equiv \sigma/N_c =
\delta \sigma / {\mu}$, which characterizes the granularity of the
order flow.  A non-dimensional scale parameter based on tick size is
constructed by dividing the tick size $dp$ by the characteristic
price, i.e. $dp/p_c = {\alpha} dp /{\mu}$.  The usefulness of this is
that the properties of the model only depend on the two
non-dimensional parameters, $\epsilon$ and $dp/p_c$: Any variations of
the parameters $\mu$, $\alpha$, and $\delta$ that keep these two
non-dimensional parameters constant gives exactly the same market
properties. One of the interesting results that emerges from analysis
of the model is that the effect of the granularity parameter
$\epsilon$ is generally much more important than the tick size
$dp/p_c$. For a more detailed discussion, see
reference~\cite{Smith03}.

While $a(t)$ and $b(t)$ make random walks, the increments of their
random walks are strongly anti-correlated.  This is a good example of
how the properties of this model are not simple to understand.  One
might naively think that under IID Poisson order flow, price
increments should also be IID.  However, due to the coupling of
boundary conditions for the buy market order/sell limit order process
to those of the sell market order/buy limit order process, this is not
the case.  Because of the fact that supply and demand tend to build as
one moves away from the center of the book, price reversals are more
common than price changes in the same direction.  As a result, the
price increments generated by this model are more anti-correlated than
those of real price series.  This has an interesting consequence: If
we add the assumption of market efficiency, and assume that real price
increments must be white, it implies that real order flow should be
positively autocorrelated in order to compensate for the
anticorrelations induced by the continuous double auction.  This has
indeed been observed to be the case~\cite{Bouchaud03.2,Lillo03.2}.

This is of course also a criticism of the model, since it implies a
lack of arbitrage efficiency.  However, we wish to stress that we make
no claims that this model explains everything about the market; just
that it explains a few things fairly well.

\subsection{The London Stock Exchange (LSE) data set}
\label{datasetDesc}

The London Stock Exchange is composed of two parts, the electronic
open limit order book, and the non-electronic upstairs market, which
is used to facilitate large block trades.  During the time period of
our dataset 40\% to 50\% of total volume was routed through the
electronic order book and the rest through the upstairs market. It
is believed that the limit order book is the dominant price formation
mechanism of the London Stock Exchange: about 75\% of upstairs trades
happen between the current best prices in the order
book~\cite{LSEbulletin}. Our analysis involves only the data from the
electronic order book.  We chose this data set to study because we
have a complete record of every action taken by every participating
institution, allowing us to measure the order flows and cancellations
and estimate all of the necessary parameters of our model.

We used data from the time period August 1st 1998 - April 30th 2000,
which includes a total of 434 trading days and roughly six million
events.  We chose 11 stocks each having the property that the number
of total number of events exceeds 300,000 and was never less than 80
on any given day.  Some statistics about the order flow for each stock
are given in table~\ref{summaryStats}.
 
\begin{table*}[!bthp]
\begin{center}
\begin{tabular}{|l| c c c c c c c c |} 
\hline
stock & num. events & average & limit & market & deletions & eff. limit & eff. market & \# days\\
ticker & (1000s) & (per day) & (1000s) & (1000s) & (1000s) & (shares) & (shares) & \\
\hline
AZN  & 608 & 1405 &  292 & 128 & 188 &  4,967   & 4,921 & 429 \\
BARC & 571 & 1318 &  271 & 128 & 172 &  7,370   & 6,406 & 433 \\
CW.  & 511 & 1184 &  244 & 134 & 134 &  12,671  & 11,151& 432 \\
GLXO & 814 & 1885 &  390 & 200 & 225 &   8,927  & 6,573 & 434 \\
LLOY & 644 & 1485 &  302 & 184 & 159 &  13,846  & 11,376& 434 \\
ORA  & 314 & 884  &  153 & 57  & 104 &  12,097  & 11,690& 432 \\
PRU  & 422 & 978  &  201 & 94  & 127 &  9,502   & 8,597 & 354 \\
RTR  & 408 & 951  &  195 & 100 & 112 &  16,433  & 9,965 & 431 \\
SB.  & 665 & 1526 &  319 & 176 & 170 &  13,589  & 12,157& 426 \\
SHEL & 592 & 1367 &  277 & 159 & 156 &  44,165  & 30,133& 429 \\
VOD  & 940 & 2161 &  437 & 296 & 207 &  89,550  & 71,121& 434 \\
\hline
\end{tabular}
\end{center}
\caption{\small Summary statistics for stocks in the dataset. Fields
  from left to right: stock ticker symbol, total number of events
  (effective market orders + effective limit orders + order
  cancellations) in thousands, average number of events in a trading
  day, number of effective limit orders in thousands, number of
  effective market orders in thousands, number of order deletions in
  thousands, average limit order size in shares, average market order
  size in shares, number of trading days in the sample.}
\label{summaryStats}
\end{table*}

The trading day of the LSE starts at 7:50 with a roughly 10 minute
long opening auction period (during the later part of the dataset the
auction end time varies randomly by 30 seconds).  During this time
orders accumulate without transactions; then a clearing price for the
opening auction is calculated, and all opening transactions take place
at this price.  Following the opening at 8:00 the market runs
continuously, with orders matched according to price and time
priority, until the market closes at 16:30.  In the earlier part of
the dataset, until September 22nd 1999, the market opening hour was
9:00.  During the period we study there have been some minor
modifications of the opening auction mechanism, but since we discard
the opening auction data anyway this is not relevant.

Some stocks in our sample (VOD for example) have stock price splits
and tick price changes during the period of our sample.  We take
splits into account by transforming stock sizes and prices to
pre-split values.  In any case, since all measured quantities are in
logarithmic units, of the form $\log(p_1) - \log(p_2)$, the absolute
price scale drops out.  Our theory predicts that the tick size should
change some of the quantities of interest, such as the bid-ask spread,
but the predicted changes are small enough in comparison with the
effect of other parameters that we simply ignore them (and base our
predictions on the limit where the tick size is zero).  Since
granularity is much more important than tick size, this seems to be a
good approximation.

\subsection{Measurement of model parameters}
\label{param_measurement}
Our goal is to compare the predictions of the model with real data.
The parameters of the model are stated in terms of order arrival
rates, cancellation rate, order size, and tick size.  We choose an
appropriate time interval and measure the parameters over that
interval, and then compare to the properties of the market over that
same interval.

Reconstructing the limit order book on a moment-by-moment basis makes
it clear that the properties of the market tend to be relatively
stationary during each day, changing more dramatically at the
beginning and at the end of day.  It is therefore natural to measure
each parameter for each stock on each day.  Since the model does not
take the opening auction into account, we simply neglect orders
leading up to the opening auction, and base all our measurements on
the remaining part of the trading day, when the auction is continuous.
Averaging daily parameters, rather than computing the parameters
directly across the whole period, has the important advantage in
computing volatility, of neglecting the effect of overnight price
movements, which our model does not attempt to explain.

In order to treat simply and in a unified manner the diverse types of
orders traders can submit in a real market (for example, crossing
limit orders, market orders with limiting price, `fill-or-kill,
execute \& eliminate) we use redefinitions based on whether an order
results in an immediate transaction, in which case we call it an
\emph{effective market order}, or whether it leaves a limit order
sitting in the book, in which case we call it an \emph{effective limit
  order}.  Marketable limit orders (also called crossing limit orders)
are limit orders that cross the opposing best price, and so result in
at least a partial transaction.  The portion of the order that results
in an immediate transaction is counted as a effective market order,
while the non-transacted part (if any) is counted as a effective limit
order. Orders that do not result in a transaction and do not leave a
limit order in the book, such as for example, failed fill-or-kill
orders, are ignored altogether.  These have no affect on prices, and
in any case, make up only a very small fraction of the order flow,
typically less than 1\%.  Note that we drop the term ``effective", so
that e.g. ``market order" means ``effective market order".

A limit order can be removed from the book for many reasons,
e.g. because the agent changes her mind, because a time specified when
the order was placed has been reached, or because of the
institutionally-mandated $30$ day limit on order duration.  We will
lump all of these together, and simply refer to them as
``cancellations".

Our measure of time is based on the number of events, i.e., the time
elapsed during a given period is just the total number of events,
including effective market order placements, effective limit order
placements, and cancellations.  We call this \emph{event time}.  Price
intervals are computed as the difference in the logarithm of prices,
which is consistent with the model, in which all price intervals are
assumed to be logarithmic in order to assure prices are always
positive.

We measure the average value of the five parameters of the model,
$\mu$, $\alpha$, $\delta$, $\sigma$, and $dp$ for each day.  This has
the advantage that it allows us to skip over the opening auction, but
is not essential for this analysis.  $\mu$, $\sigma$, and $dp$ are
straightforward to measure, but there are problems in measuring
$\alpha$ and $\delta$ that must be understood in order to properly
interpret our results.

The parameter $\mu_t$, which characterizes the average market order
arrival rate on day $t$, is straightforward to measure.  It is just
the ratio of the number of shares of effective market orders (for both
buy and sell orders) to the number of events during the trading
day. Similarly, $\sigma_t$ is the average limit order
size\footnote{The model assumes that the average size of limit orders
  and market orders is the same.  For the real data this is not
  strictly true, though as seen in Table \ref{summaryStats}, it is a
  good approximation to within about $20\%$.  For the purposes of the
  analysis we use the limit order size as the measure because for
  theoretical reasons we think this is more important than the market
  order size, but because the two are approximately the same, this
  will not make a significant difference in the results.}  in shares
for that day.

Measuring the cancellation rate $\delta_t$ and the limit order rate
density $\alpha_t$ is more complicated, due to the highly simplified
assumptions we have made for the model.  In contrast to our assumption
of a constant density for placement of limit orders across the entire
logarithmic price axis, real limit order placement is highly
concentrated near the best prices (roughly $2/3$ of all orders are
placed at inside of the best prices), with a density that falls off as
a power law as a function of the distance $\Delta$ from the best
prices~\cite{Bouchaud02,Zovko02}.  In addition, we have assumed a
constant cancellation rate, whereas in reality orders placed near the
best prices tend to be cancelled much faster than orders placed far
from the best prices.  We cope with these problems as described below.

In order to estimate the limit order rate density for day $t$,
$\alpha_t$, we make an empirical estimate of the distribution of the
relative price for effective limit order placement on each day.  For
buy orders we define the relative price as $\Delta = m - p$, where $p$
is the logarithm of the limit price and $m$ is the logarithm of the
midquote price. Similarly for sell orders, $\Delta = p - m$.  We then
somewhat arbitrarily choose $Q_t^{\mbox{\small{lower}}}$ as the 2
percentile of the density of $\Delta$ corresponding to the limit
orders arriving on day $t$, and $Q_t^{\mbox{\small{upper}}}$ as the 60
percentile of $\Delta$.  Assuming constant density within this range,
we calculate $\alpha_t$ as $\alpha_t = L/(Q_t^{\mbox{\small{upper}}} -
Q_t^{\mbox{\small{lower}}})$ where $L$ is the total number of shares
of effective limit orders within the price interval
$(Q_t^{\mbox{\small{lower}}}, Q_t^{\mbox{\small{upper}}})$ on day $t$.
  These choices are made in a compromise to include as much data as
  possible for statistical stability, but not so much as to include
  orders that are unlikely to ever be executed, and therefore unlikely
  to have any effect on prices.

\begin{figure*}[hbt]
\begin{center} 
   \includegraphics[scale=0.6]{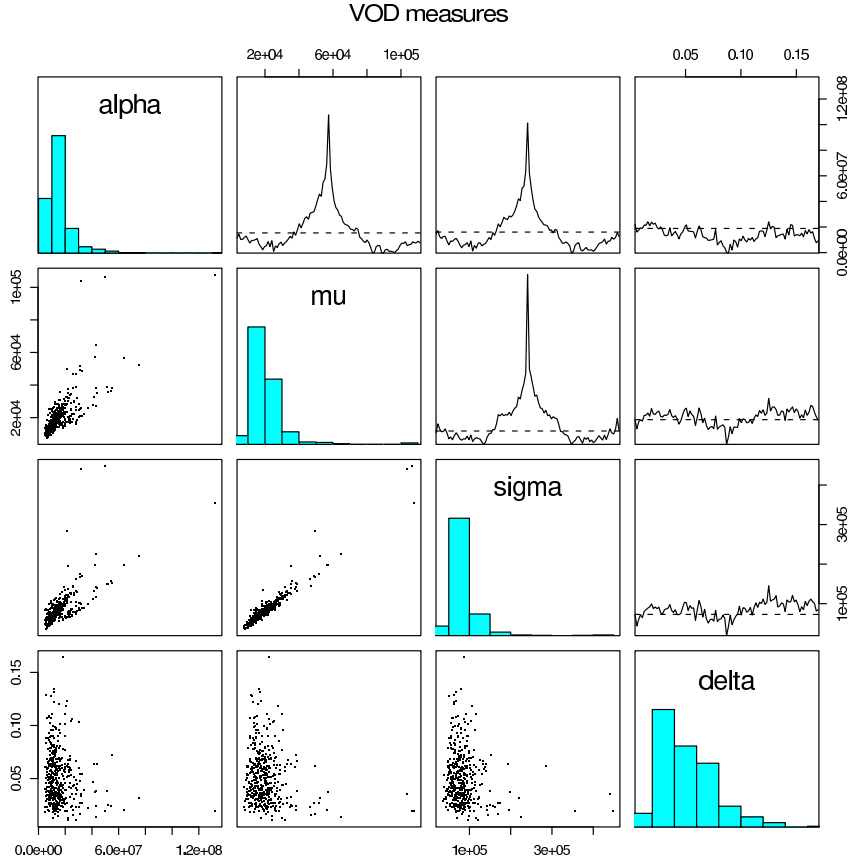} 
   \caption{Density estimations and cross correlations for 
        Vodafone between the four model parameter measures.  On the
        diagonal we present the histogram of the corresponding
        parameter.  Upper off-diagonal plots are the time cross
        correlation.  We see that $\delta$ is uncorrelated with other
        measures, while the other three are quite correlated although
        without any noticeable lead-lag effects.  The lower
        off-diagonal plots are scatter plots between the
        parameters. $\mu$ and $\alpha$ are particularly strongly
        correlated; fortunately, for the prediction of the spread
        their hgratio is the most important quantity, and this
        correlation largely cancels out.}
\label{VODpairs}
\end{center}
\end{figure*}

Similarly, to cope with the fact that in reality the average
cancellation rate $\delta$ decreases \cite{Bouchaud02} with the
relative price $\Delta$, whereas in the model $\delta$ is assumed to
be constant, we base our estimate for $\delta$ only on canceled limit
orders within the range of the same relative price boundaries
$(Q_t^{\mbox{\small{lower}}}, Q_t^{\mbox{\small{upper}}})$ defined
above.  We do this to be consistent in our choice of which orders are
assumed to contribute significantly to price formation (orders closer
to the best prices contribute more than orders that are further away).
We then measure $\delta_t$, the cancellation rate on day $t$, as the
inverse of the average lifetime of a canceled limit order in the above
price range.  Lifetime is measured in terms of number of events
happening between the introduction of the order and its subsequent
cancellation.  Some simple diagnostics of the parameter estimates are
presented in Fig.~\ref{VODpairs}.

\subsection{Measuring the price diffusion rate}

\begin{figure}[tbh]
   \begin{center} 
   \includegraphics[scale=0.4, angle=90]{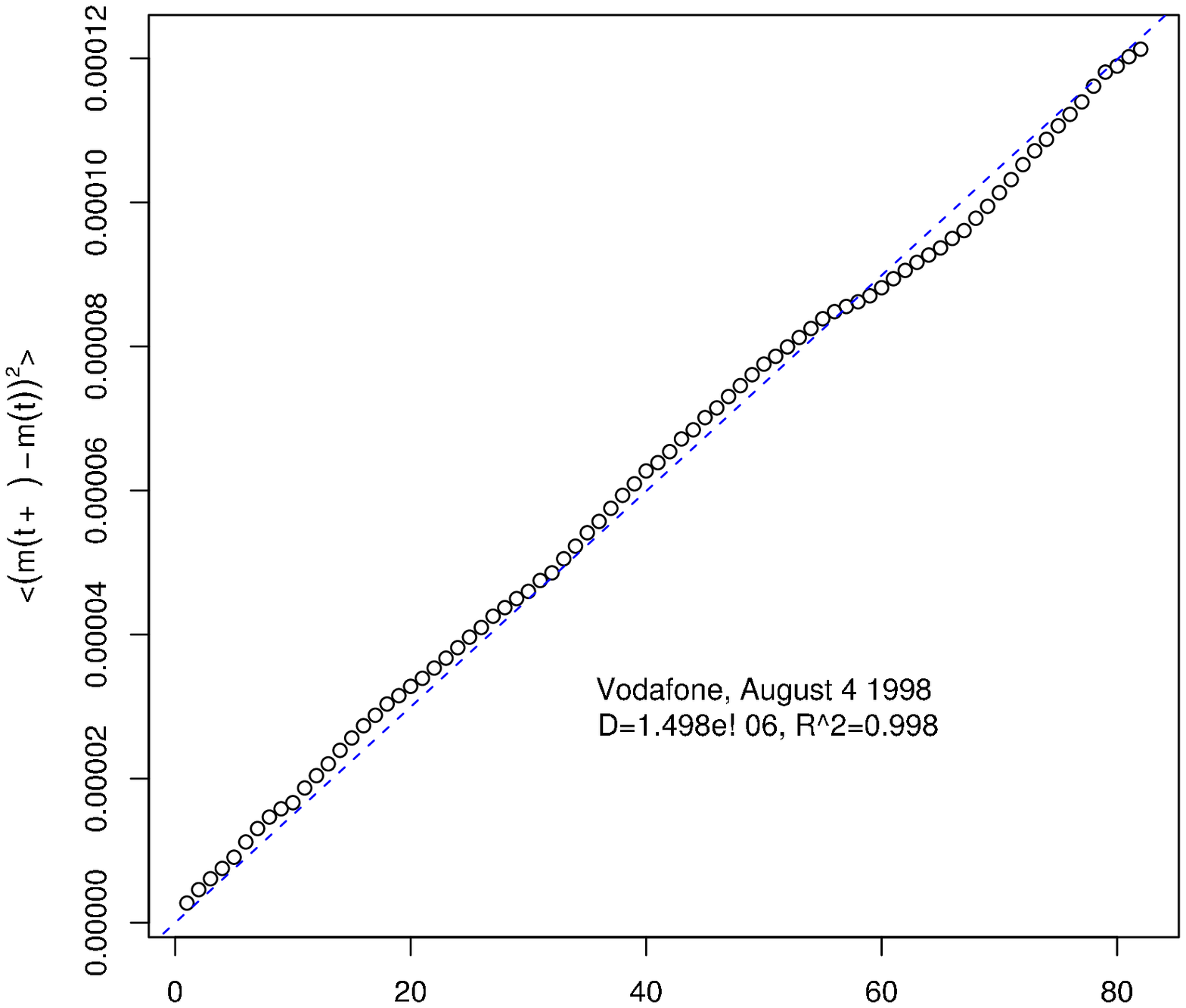}
   \caption{Illustration of the procedure for measuring the price
     diffusion rate for Vodafone (VOD) on August 4th, 1998.  On the
     $x$ axis we plot the time $\tau$ in units of ticks, and on the
     $y$ axis the variance of mid-price diffusion $V(\tau)$.
     According to the hypothesis that mid-price diffusion is an
     uncorrelated Gaussian random walk, the plot should obey $V(\tau)
     = D \tau$.  To cope with the fact that points with larger values
     of $\tau$ have fewer independent intervals and are less
     statistically significant, we use a weighted regression to
     compute the slope $D$.}
   \label{volEst} 
   \end{center}
\end{figure}

The measurement of the price diffusion rate requires some discussion.
We measure the intraday price diffusion by computing the variance
$V(\tau)$ of $m(i - \tau) - m(i)$, averaged over different intraday
events $i$. Here an event is anything that changes the midpoint price
$m$.  If we assume that the events are asymptotically IID, then the
estimated slope of the variance plot is the diffusion rate $D_t$ for
day $t$.  To compute this we regress $V(\tau)$ against $\tau$, using
the assumption $V(\tau) = D_t \tau$.  We use an ordinary least squares
regression to estimate $D_t$, weighting each value of $\tau$ by the
square root of the number of independent observations.  An example of
this procedure is given in Fig.~\ref{volEst}.

\begin{figure}[bht]
   \begin{center} 
   \includegraphics[scale=0.4, angle=0]{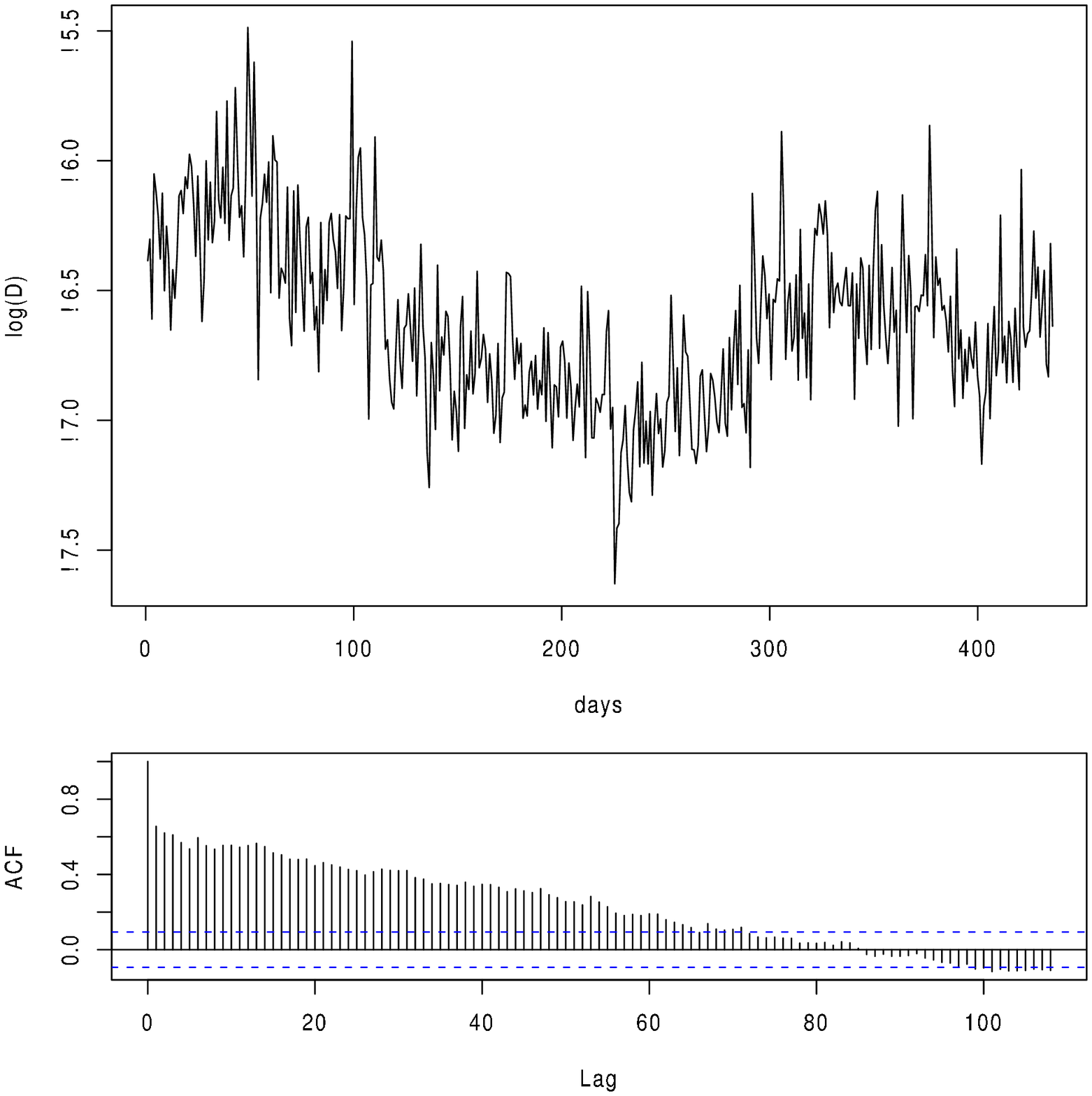}
   \caption{Time series (top) and autocorrelation function (bottom)
     for daily price diffusion rate $D_t$ for Vodafone.  Because of
     long-memory effects and the short length of the series, the
     long-lag coefficients are poorly determined; the figure is just
     to demonstrate that the correlations are quite large.}
      \label{diffTs} 
   \end{center}
\end{figure}

One must bear in mind that the price diffusion rate from
day to day has substantial correlations, as illustrated in
Fig.~\ref{diffTs}.

\subsection{Estimating the errors for the regressions}
\label{regressionErrors}

The error bars presented in the text are based on a bootstrapping
method.  We are driven to use this method for two reasons: First, the
spread, price diffusion rates, and parameters are highly
cross-correlated between stocks, and second, because order flow
variables, spread, and price diffusion rates all have slowly decaying
positive autocorrelation functions.  Indeed, it has recently been
shown that order sign, order volume and liquidity as reflected by
volume at the best price, are long-memory processes
\cite{Bouchaud03.2,Lillo03.2}.  These effects complicate the
statistical analysis, and make the assignment of error bars difficult.

\begin{figure*}[hbtp]
\begin{center}
\includegraphics[scale=0.55]{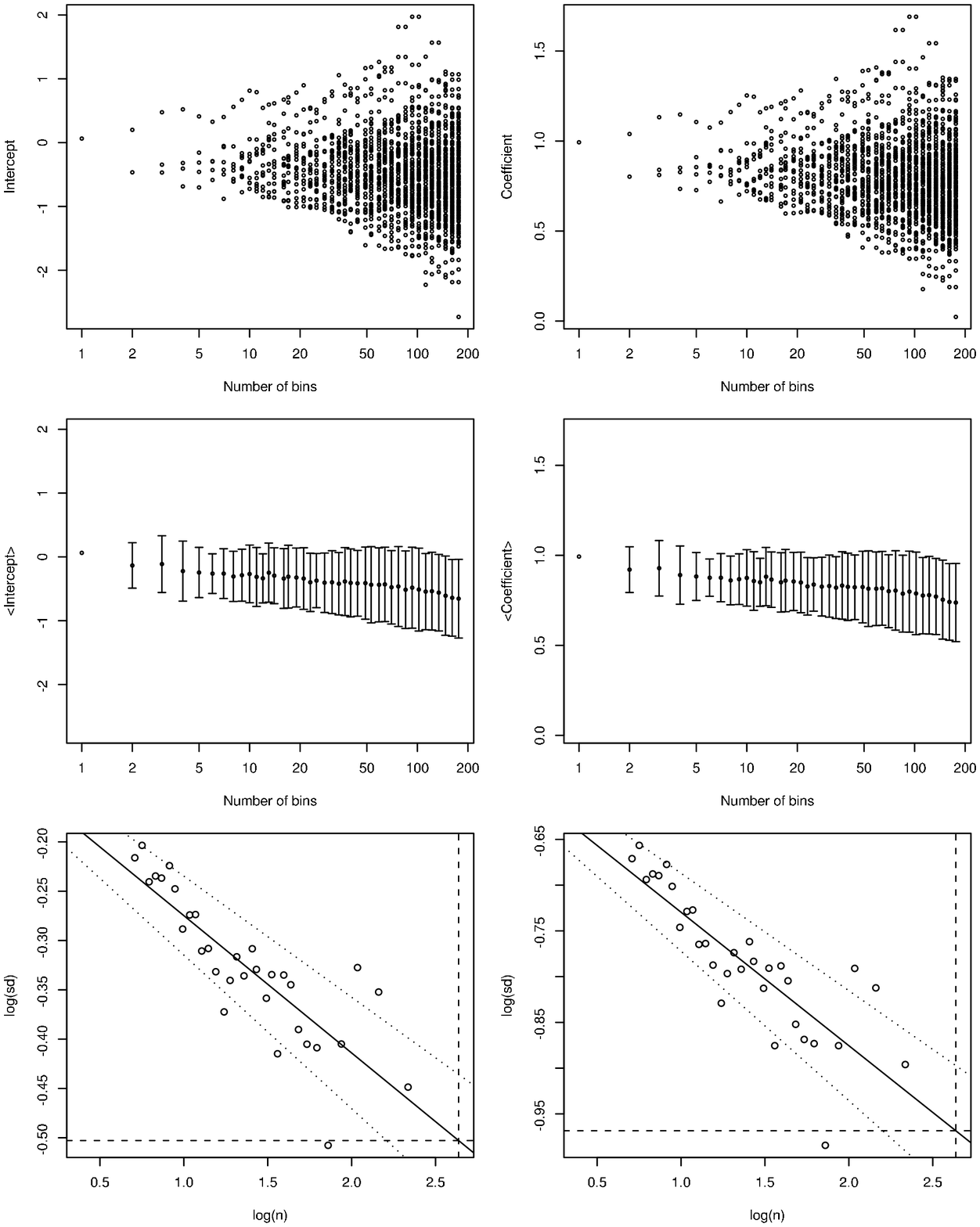}
\caption{Subsample analysis of regression of predicted
    vs. actual spread.  To get a better
    feeling for the true errors in this estimation (as opposed to
    standard errors which are certainly too small), we divide the data
    into subsamples (using the same temporal period for each stock)
    and apply the regression to each subsample. (a) (top left) shows
    the results for the intercept, and (b) (top right) shows the
    results for the slope.  In both cases we see that progressing from
    right to left, as the subsamples increase in size, the estimates
    become tighter.  (c) and (d) (next row) shows the mean and
    standard deviation for the intercept and slope.  We observe a
    systematic tendency for the mean to increase as the number of bins
    decreases.  (e) and (f) show the logarithm of the standard
    deviations of the estimates against $\log n$, the number of each
    points in the subsample.  The line is a regression based on
    binnings ranging from $m=N$ to $m = 10$ (lower values of $m$ tend
    to produce unreliable standard deviations).  The estimated error
    bar is obtained by extrapolating to $n=N$.  To test the accuracy
    of the error bar, the dashed lines are one standard deviation
    variations on the regression, whose intercepts with the $n=N$
    vertical line produce high and low estimates.}
\label{spreadBootstrap}
\end{center}
\end{figure*}

\begin{figure*}[hbtp]
   \begin{center}
         \includegraphics[scale=0.6]{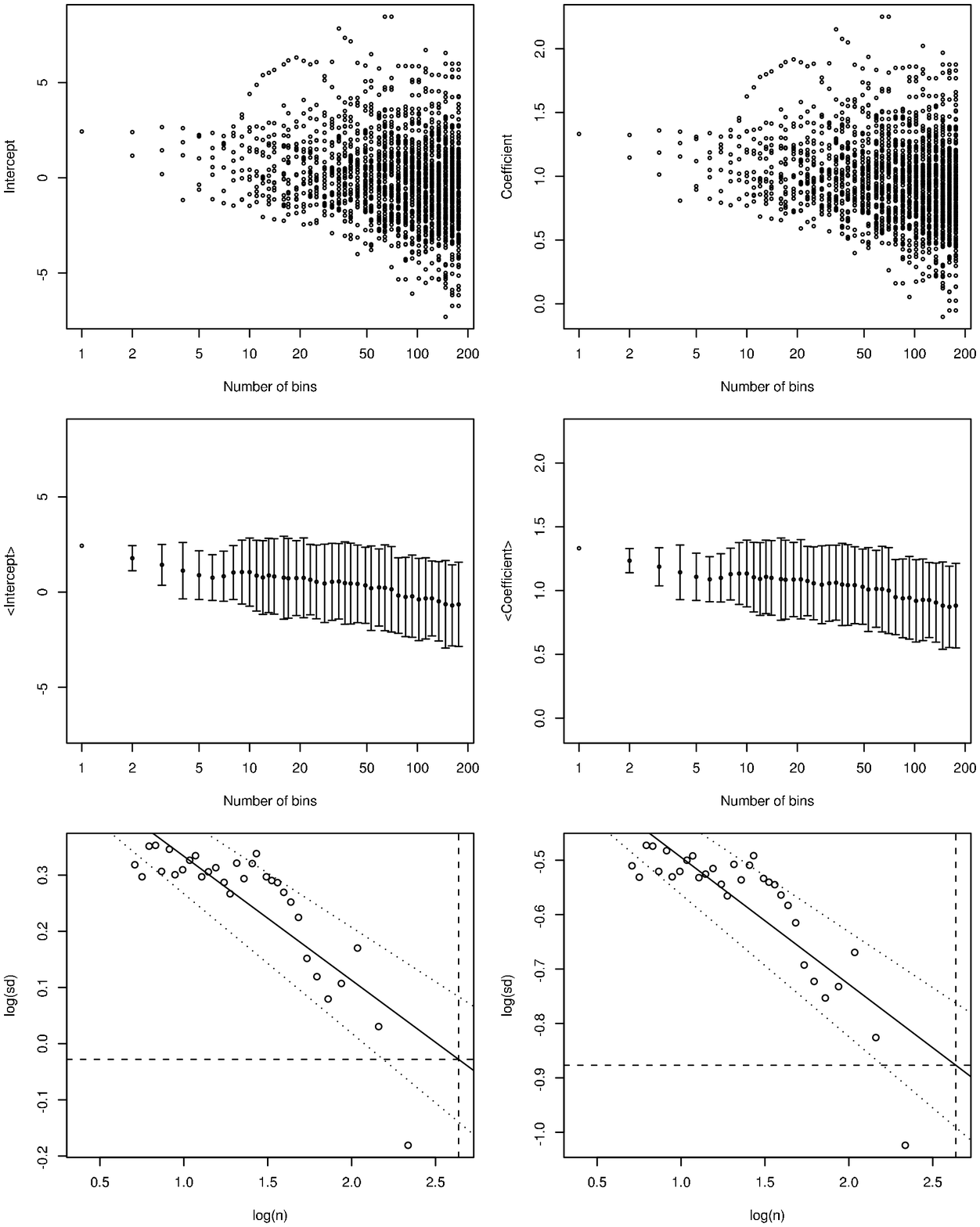} 
         \caption{Subsample analysis of regression of predicted
    vs. actual price diffusion (see Fig. 6), similar to
    the previous figure for the spread.  The scaling of the errors is
    much less regular than it is for the spread, so the error bars are
    less accurate.}
\label{diffusionBootstrap}
\end{center}
\end{figure*}

The method we use is inspired by the variance plot method described in
Beran~\cite{beran}, Section 4.4.  We divide the sample into blocks,
apply the regression to each block, and then study the scaling of the
deviation in the results as the blocks are made longer to coincide
with the full sample.  We divide the $N$ daily data points for each
stock into $m$ disjoint blocks, each containing $n$ adjacent days, so
that $n \approx N/m$.  We use the same partition for each stock, so
that corresponding blocks for each stock are contemporaneous.  We
perform an independent regression on each of the $m$ blocks, and
calculate the mean $M_m$ and standard deviation $\sigma_m$ of the $m$
slope parameters $A_i$ and intercept parameters $B_i$, $i = 1, \ldots,
m$.  We then vary $m$ and study the scaling as shown in
Figs.~\ref{spreadBootstrap}~and~\ref{diffusionBootstrap}.

Figs.~\ref{spreadBootstrap}(a) and (b) illustrate this procedure for
the spread, and Figs.~\ref{diffusionBootstrap}(a) and (b) illustrate
this for the price diffusion rate.  Similarly, panels (c) and (d) in
each figure show the mean and standard deviation for the intercept and
slope as a function of the number of bins.  As expected, the standard
deviations of the estimates decreases as $n$ increases.  The logarithm
of the standard deviation for the intercept and slope as a function of
$\log n$ is shown in panels (e) and (f). For IID normally distributed
data we expect a line with slope $\gamma = -1/2$; instead we observe
$\gamma > -1/2$.  For example for the spread $\gamma \approx
-0.19$. $|\gamma| < 1/2$ is an indication that this is a long
memory process; see the discussion in Section (\ref{marketImpact}).

This method can be used to extrapolate the error for $m = 1$, i.e. the
full sample.  This is illustrated in panels (e) and (f) in each
figure.  The inaccuracy in these error bars is evident in the
unevenness of the scaling.  This is particularly true for the price
diffusion rate.  To get a feeling for the accuracy of the error bars,
we estimate the standard deviation for the scaling regression assuming
standard error, and repeat the extrapolation for the one standard
deviation positive and negative deviations of the regression lines, as
shown in panels (e) and (f) of Figs.~\ref{spreadBootstrap} and
\ref{diffusionBootstrap} The results are summarized in
Table~\ref{errorBars}.

\begin{table*}[hbtp]
\begin{center}
\begin{tabular}{|l| c c c c c |} \hline
regression & estimated & standard & bootstrap & low & high \\   
\hline
spread intercept & 0.06 & 0.21 & 0.29 & 0.25 & 0.33 \\
spread slope & 0.99 & 0.08 & 0.10 & 0.09 & 0.11 \\
diffusion intercept & 2.43 & 1.22 & 1.76 & 1.57 & 1.97 \\
diffusion slope & 1.33 & 0.19 & 0.25 & 0.23 & 0.29 \\
\hline
\end{tabular}
\end{center}
\caption{A summary of the bootstrap error analysis described
    in the text.  The columns are (left to right) the estimated value
    of the parameter, the standard error from the cross sectional
    regression in Fig.~6, the one standard deviation
    error bar estimated by the bootstrapping method, and the one standard
    deviation low and high values for the extrapolation, as shown in
    Figs.~\ref{spreadBootstrap}(e-f) and~\ref{diffusionBootstrap}(e-f).}
\label{errorBars}
\end{table*}

One of the effects that is evident in Figs.~\ref{spreadBootstrap}(c-d)
and \ref{diffusionBootstrap}(c-d) is that the slope coefficients tend
to decrease as $m$ increases.  We believe this is due to the
autocorrelation bias discussed in Section~(\ref{longVsCross}).

\subsection{Longitudinal vs. cross-sectional tests}
\label{longVsCross}
It is possible to test this model either longitudinally (across
different time intervals for a given stock) or cross-sectionally
(across different stocks over the same time period).  We have applied
tests of both types, but due to the very strong autocorrelations of
the order flow rates, spread, and price diffusion rates, there are
difficulties in getting a clean test of the model longitudinally.  In
this section we discuss these problems, and discuss some of our
results on the longitudinal tests.

\emph{A priori} we would expect to do a better job making
cross-sectional rather than longitudinal predictions.  Indeed, it is
not clear that this model should predict anything at all about
longitudinal variations.  To see why, imagine that the assumptions of
the model are satisfied perfectly, and suppose that the five
parameters of the order flow process ($\mu$, $\alpha$, etc.) for a
given stock are fixed in time.  Then the only daily variations we
would observe in testing the model would be due to sample errors in
the estimation process.  Even though the assumptions are satisfied
perfectly, we would find no correlation between predicted and actual
values.  To observe such a correlation requires real variations in the
parameters of the order flow process.  There are also possible
problems with relaxation times: If a parameter is suddenly changed,
according to the model it takes the system time to reach a new steady
state behavior.  There are two characteristic times in the model:
$\sigma/\mu$, which is the characteristic time for removal of limit
orders by market orders, and $1/\delta$, which is is the
characteristic time for spontaneous removal of limit orders.  For the
data here it appears that $\sigma/\mu$ is typically less than a
minute, whereas $1/\delta$ ranges from a few minutes to a few
hours. Thus, $1/\delta$ is the slowest relaxation time, and in some
cases at least it is potentially problematic for a daily analysis.  In
addition, there is the very significant problem that real order flows
are strongly autocorrelated, discussed below.

Cross-sectionally, in contrast, we expect \emph{a priori} that
different stocks should have different parameters.  There are likely
to be larger variations in the parameters between stocks than in the
parameters for a given stock at different times.  In addition, for a
cross-sectional analysis there are no problems with relaxation times,
and in any case averaging over longer periods of time reduces the
sampling error.  Thus cross-sectional analysis is expected to be more
promising and more reliable.

As noted, for the daily analysis, and even for cross-sectional
analysis over long periods of time, there are problems caused by the
long range autocorrelations of real order flow, spreads, and price
diffusion rates.  Autocorrelations can remain strongly positive on the
order of 50 days.  This creates problems in performing the regression,
and can result in a systematic bias in the estimated parameters.  It
causes severe systematic biases and interpretation problems for a
daily analysis.

To produce estimates of the average values of the parameters and of
the price diffusion and spread across the full 21 month period for the
cross-sectional regressions, we have used the event-weighted average
of the daily values.  The alternative would have been to repeat the
measurements as done for the daily data on a 21 month rather than a
daily time-scale. However, this latter approach would run into
problems because of the opening auction, which is not treated by our
model.  There are price changes driven by the orders received during
the opening auction, and if we measured price diffusion across the
full period we would be including these as well as the intra-day price
movements.  As a simple solution to this problem we use an
event-weighted 21 month average of daily values to compute values for
each of the order flow parameters, and then make predictions for each
stock based on the average values.  The weighting is done by the
number of events in a day, which for simple quantities such as the
market impact rate reduces to something that is equivalent to applying
the analysis over the full period.  Similarly, to get the 21 month
average of the spread and price diffusion we simply compute an
event-weighted average of their daily values.  We have tried several
variations on this procedure and the differences appear to be
inconsequential.

When we perform longitudinal regressions at a daily time-scale we get
values for the slope coefficient of the regressions that are less than
one, often by a statistically significant amount.  We believe this is
caused by the strong autocorrelation.  For example, consider a time
series process of the form
\begin{equation}
y_t = ax_t + \rho y_{t-1} + n_t
\end{equation}
where $n_t$ is an IID noise process.  In case $x_t$ are i.i.d.,
regressing $y_t$ against $x_t$ will result in coefficients that are
systematically too small, due to the fact that the $y_{t-1}$ term
damps the response of $y_t$ to changes in $x_t$.  Of course, one can
fix this in the simple example above by simply including $y_{t-1}$ in
the regression~\cite{greene}.  For the real data, however, the
autocorrelation structure is more complicated -- indeed we believe it
is a long-memory process -- which is not well modeled by an AR process
in the above form.  Without finding a proper characterization of the
autocorrelation structure, we are likely to make errors in estimating
the dependence of the predicted and actual values.  This is borne out
in the error analysis presented in Section~(\ref{regressionErrors}),
where we see that as we break the data into shorter subsamples, the
estimated slope coefficients systematically decrease for the spread
and the price diffusion.

\begin{figure*}[htpb]
   \begin{center} 
   \includegraphics[scale=0.6]{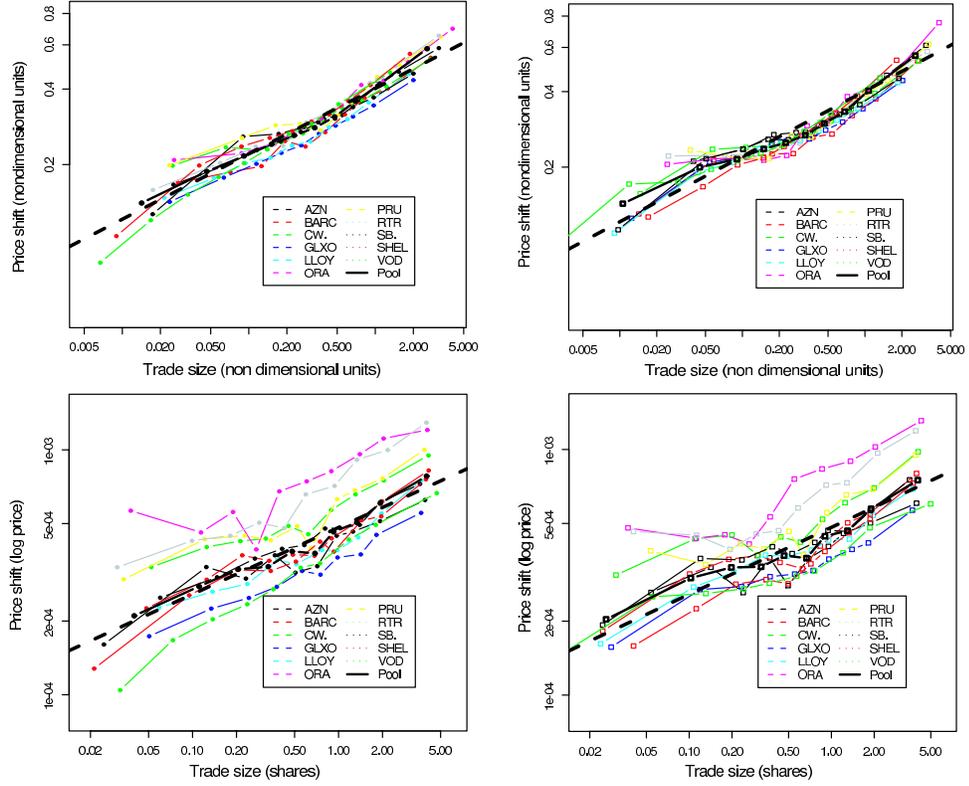}
   \caption{\small The average market impact vs. order size plotted on
     log-log scale.  The upper left and right panels show buy and sell
     orders in non-dimensional coordinates; the fitted line has slope
     $\beta = 0.26 \pm 0.02$ for buy orders and $\beta = 0.23 \pm
     0.02$ for sell orders.  In contrast, the lower panels show the
     same thing in dimensional units, using British pounds to measure
     order size.  Though the exponents are similar, the scatter
     between different stocks is much greater.}
  \label{loglog}
  \end{center}
\end{figure*}

If we fit a function of the form $\phi(\omega)= K \omega^\beta$ to the
market impact curve, we get $\beta = 0.26 \pm 0.02$ for buy orders and
$\beta = 0.23 \pm 0.02$ for sell orders, as shown in
Fig.~\ref{loglog}.  The functional form of the market impact we
observe here is not in agreement with a recent theory by Gabaix et
al.~\cite{Gabaix03}, which predicts $\beta = 0.5$.  While the error
bars given are standard errors, and are certainly too optimistic, it
is nontheless quite clear that the data are inconsistent with $\beta =
1/2$, as discussed in Ref.~\cite{Farmer03}.  This relates to an
interesting debate: The theory for average market impact put forth by
Gabaix et al. follows traditional thinking in economics, and
postulates that agents optimize their behavior to maximize profits,
while the theory we test here assumes that they behave randomly, and
that the form of the average market impact function is dictated by the
statistical mechanics of price formation.

\subsection{Market impact}
\label{marketImpact}

The market impact function is closely related to the more familiar
notions of supply and demand.  We have chosen to measure average
market impact in this paper rather than average relative supply and
demand for reasons of convenience. Measuring the average relative
supply and demand requires reconstructing the limit order book at each
instant, which is both time consuming and error prone.  The average
market impact function, in contrast, can be measured based on a time
series of orders and best bid and ask prices.

At any instant in time the stored queue of sell limit orders reveals
the quantity available for sale at each price, thus showing the
supply, and the stored buy orders similarly show the revealed
demand. The price shift caused by a market order of a given size
depends on the stored supply or demand through a moment expansion
\cite{Smith03}.  Thus, the collapse of the market impact function
reflects a corresponding property of supply and demand. Normally one
would assume that supply and demand are functions of human production
and desire; the results we have presented here suggest that on a short
timescale in financial markets their form is dictated by the dynamical
interaction of order accumulation, removal by market orders and
cancellation, and price diffusion.

\subsection{Alternative market impact collapse plots}

\begin{figure*}[tbhp]
   \begin{center} 
  \includegraphics[scale=0.7]{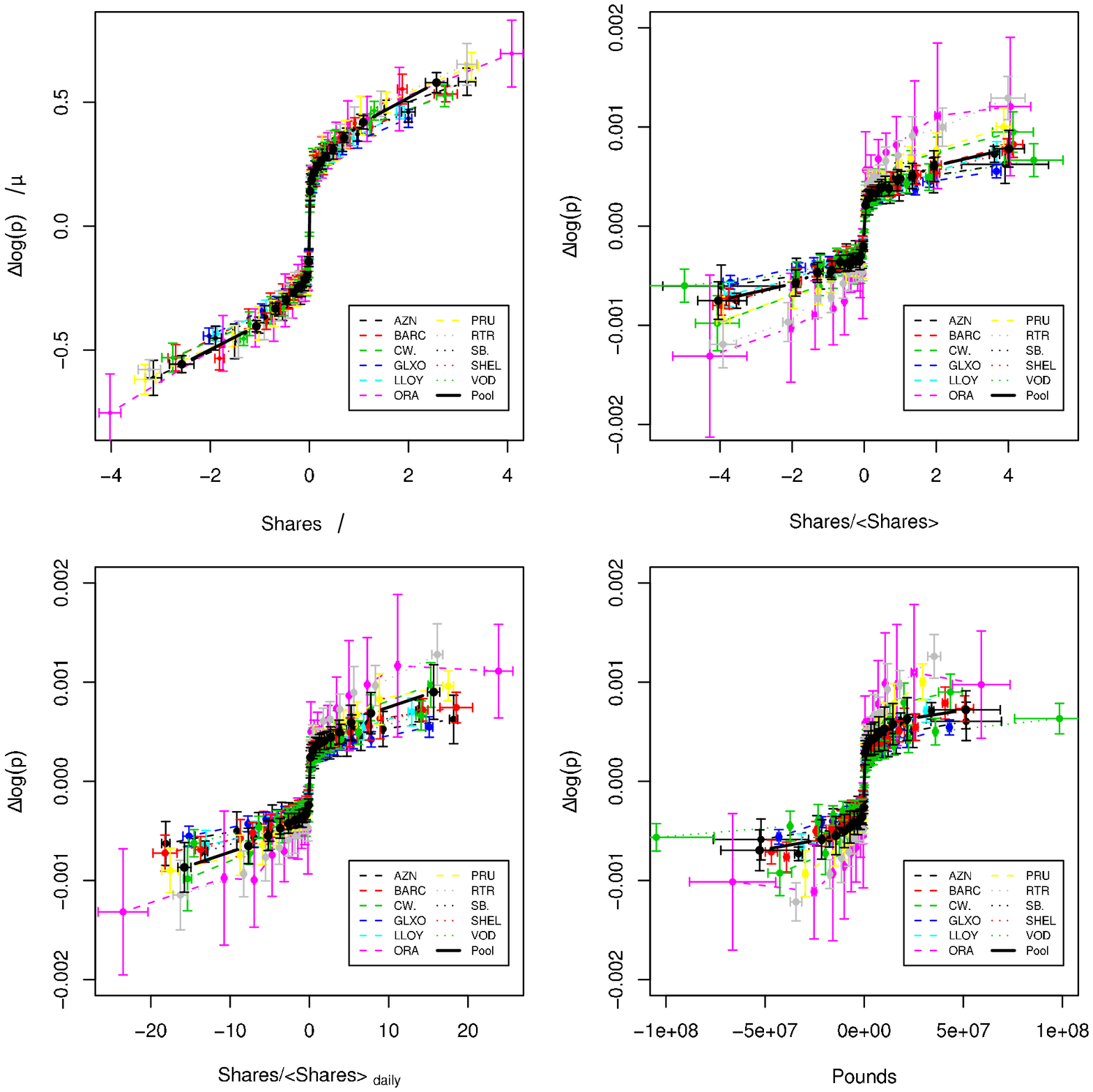}
   \caption{\textsf{Market impact collapse under 4 kinds of axis rescaling.  In each case
we plot a normalised version of the order size on the horizontal axis
vs. a (possibly normalised) average market impact $\log (p_{t+1}) - \log(p_t)$
on the vertical axis. (a) (top left) collapse using non-dimensional units
based on the model; (b) (top right) order size
is normalised by its mean value for the sample.  
(c) (bottom right) order size is normalised
the average daily volume. (d) (bottom right) Order size is multiplied
by the current best midpoint price, making the horizontal axis
the monetary value of the trade.}}
  \label{altCollapse}
  \end{center}
\end{figure*}

We have demonstrated a good collapse of the market impact using
nondimensional units.  However, in deciding what ``good'' means, one
should compare this to the best alternatives available.  We compare to
three such alternatives.  
In figure~\ref{altCollapse}, the top left
pane shows the collapse when using non-dimensional units derived from
the model (repeated from the main text).  The top right plot shows the
average market impact when we instead normalise the order size by its
sample mean.  Order size is measured in units of shares and market
impact is in log price difference.  The bottom left attempts to take
into account daily variations of trading volume, normalising the order
size by the average order size for that stock on that day.  In the
bottom right we use trade price to normalise the order sizes which are
now in monetary units (British Pounds).  We visually see that none of
the alternative rescalings comes close to the collapse we obtain when
using non-dimensional units; because of the much greater dispersion,
the error bars in each case are much larger.

\subsection{Error analysis for market impact}
\label{marketImpactErrors}
Assigning error bars to the average market impact is difficult because
the absolute price changes $\Delta p$ have a slowly decaying positive
autocorrelation function.  This may be a long-memory process, although
this is not as obvious as it is for other properties of the market,
such as the volume and sign of orders \cite{Bouchaud03.2,Lillo03.2}.
The \emph{signed} price changes $\Delta p$ have an autocorrelation
function that rapidly decays to zero, but to compute market impact we
sort the values into bins, and all the values in the bin have the same
sign.  One might have supposed that because the points entering a
given bin are not sequential in time, the correlation would be
sufficiently low that this might not be a problem.  However, the
autocorrelation is sufficiently strong that its effect is still
significant, particularly for smaller market impacts, and must be
taken into account.

\begin{figure}[h]
   \begin{center} 
        \includegraphics[scale=0.4,angle=90]{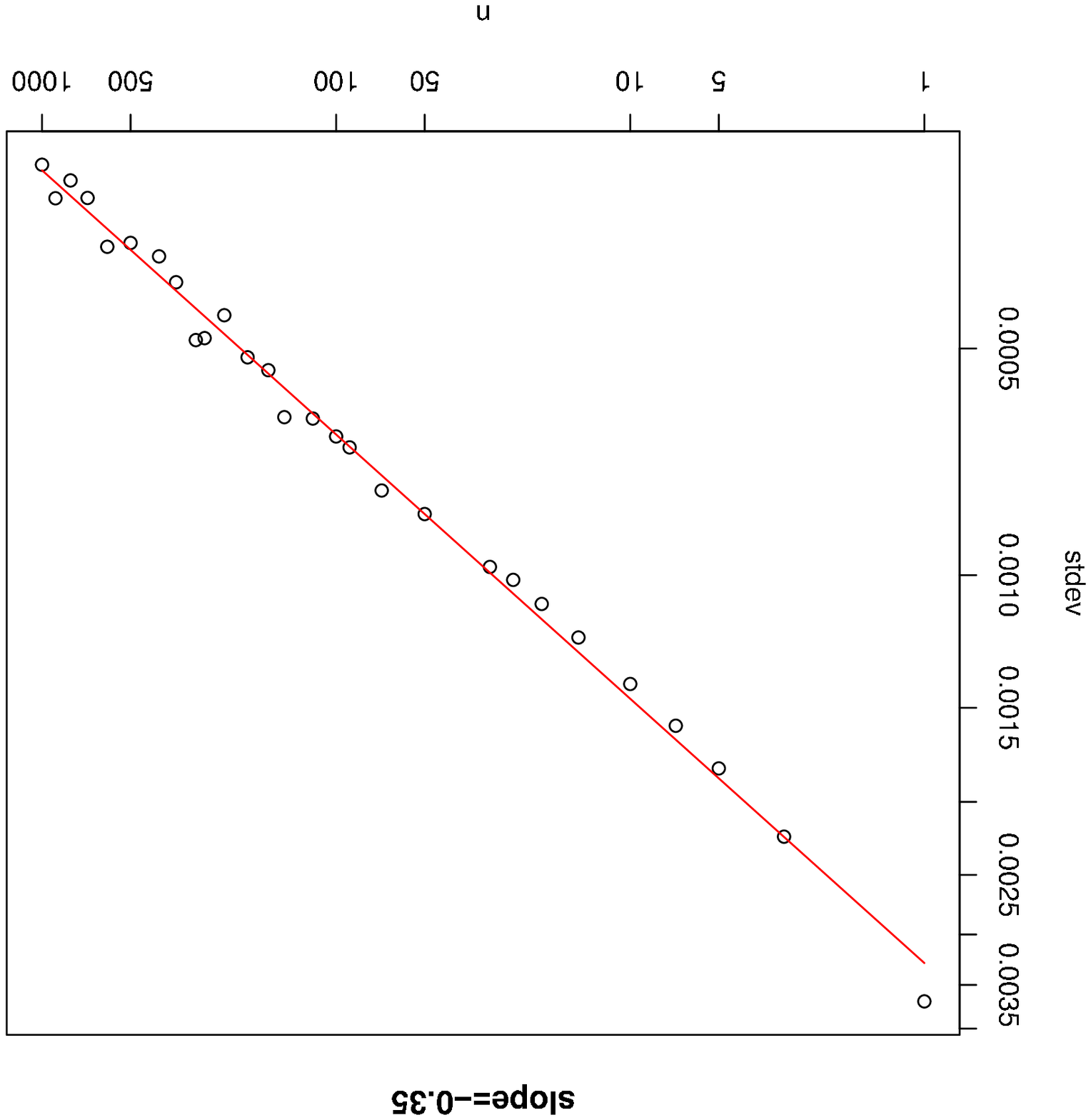} 
        \caption{The variance plot procedure used to determine error
    bars for mean market impact conditional on order size. The
    horizontal axis $n$ denotes the number of points in the $m$
    different samples, and the vertical axis is the standard deviation
    of the $m$ sample means.  We estimate the error of the full sample
    mean by extrapolating $n$ to the full sample length.}
\label{errorScaling}
\end{center}
\end{figure}

To cope with this we assign error bars to each bin using the variance
plot method described in, for example, Beran~\cite{beran}, Section
4.4.  This is a more straightforward version of the method discussed
in Section~(\ref{regressionErrors}). The sample of size $N=434$ is
divided into $m$ subsamples of $n$ points adjacent in time.  We
compute the mean for each subsample, vary $n$, and compute the
standard deviation of the means across the $m=N/n$ subsamples.  We
then make use of theorem 2.2 from Beran~\cite{beran} that states that
the error in the $n$ sample mean of a long-memory process is $\hat{e}
= \sigma{n}^{-\gamma}$, where $\gamma$ is a positive coefficient
related to the Hurst exponent and $\sigma$ is the standard
deviation. By plotting the standard deviation of the $m$ estimated
intercepts as a function of $n$ we estimate $\gamma$ and extrapolate
to $n=\mbox{sample length}$ to get an estimate of the error in the
full sample mean.  An example of an error scaling plot for one of the
bins of the market impact is given in Fig.~\ref{errorScaling}.

A central question about Fig.~\ref{marketImpactPic} is whether the
data for different stocks collapse onto a single curve, or whether
there are statistically significant idiosyncratic variations from
stock to stock.  From the results presented in
Fig.~\ref{marketImpactPic} this is not completely clear.  Most of the
stocks collapse onto the curve for the pooled data (or the pooled data
set with themselves removed). There are a few that appear to make
statistically significant variations, at least if we assume that the
mean value of the bins for different order size levels are
independent.  However, they are most definitely \emph{not}
independent, and this non-independence is difficult to model.  In any
case, the variations are always fairly small, not much larger than the
error bars.  Thus the collapse gives at least a good approximate
understanding of the market impact, even if there are some small
idiosyncratic variations it does not capture.

\subsection{Extending the model}
In the interest of full disclosure, and as a stimulus for future work,
in this section we detail the ways in which the current model does not
accurately match the data, and sketch possible improvements.  This
model was intended to describe a few average statistical properties of
the market, some of which it describes very well.  However, there are
several aspects that it does not describe well, such as the scale-free
power law properties.  This would require a more sophisticated model
of order flow, including a more realistic model of price dependence in
order placement and cancellations \cite{Bouchaud02,Zovko02},
long-memory properties \cite{Bouchaud03.2,Lillo03.2} and the
relationship of the different components of the order flow to each
other.  This is a much harder problem, and is likely to require a more
complicated model.  While this would have some advantages, it would
also have some disadvantages.

Some market properties that might profit from such an improved model
are detailed below.
\begin{itemize}
\item \emph{Price diffusion.}  The variance of real prices obeys the
  relationship $\sigma^2(\tau) = D\tau^{2H}$ to a good approximation
  for all values of $\tau$, with $H$ close to and typically a little
  greater than $0.5$.  In contrast, under Poisson order flow, due to
  the dynamics of the double continuous auction price formation
  process, prices make a strongly anti-correlated random walk, so that
  the function $\sigma^2(\tau)$ is nonlinear.  Asymptotically $H =
  0.5$, but for shorter times $H < 0.5$. Alternatively, one can
  characterize this in terms of a timescale-dependent diffusion rate
  $D(\tau)$, so that the variance of prices increases as
  $\sigma^2(\tau) = D(\tau)\tau$. Refs. \cite{Daniels03,Smith03}
  showed that the limits $\tau \rightarrow 0$ and $\tau \rightarrow
  \infty$ obey well-defined scaling relationships in terms of the
  parameters of the model.  In particular, $D(0) \sim \mu^2 \delta /
  \alpha^2 \epsilon^{-1/2}$, and $D(\infty) \sim \mu^2 \delta /
  \alpha^2 \epsilon^{1/2}$.  Interestingly, and for reasons we do not
  fully understand, the prediction $D(0)$ does a good job of matching
  the real data, as we have shown here, while $D(\infty)$ does a poor
  job. Note that it is very interesting that the double continuous
  auction produces anti-correlations in prices, even with no
  correlation in order flow.  One can turn this around: Given that
  prices are uncorrelated, there must be correlations in order flow.
  And indeed this is observed to be the case
  \cite{Bouchaud03.2,Lillo03.2}.
\item \emph{Market efficiency}.  The question of market efficiency is
  closely related to price diffusion.  The anti-correlations mentioned
  above imply a market inefficiency.  We are investigating the
  addition of ``low-intelligence'' agents to correct this problem.
\item \emph{Correlations in spread and price diffusion.} We have
  already discussed in Section~(\ref{longVsCross}) the problems that
  the autocorrelations in spread and price diffusion create for
  comparing the theory to the model on a daily scale.
\item \emph{Lack of dependence on granularity parameter.} In
  Section~(\ref{marketImpact}) we discuss the fact that the model
  predicts more variation with the granularity parameter than we
  observe.  Apparently the Poisson-based non-dimensional coordinates
  work even better than one would expect.  This suggests that there is
  some underlying simplicity in the real data that we have not fully
  captured in the model.
\end{itemize}

Although in this paper we are stressing the fact that we can make a
useful theory out of zero-intelligence agents, we are certainly not
trying to claim that intelligence doesn't play an important role in
what financial agents do.  Indeed, one of the virtues of this model is
that it provides a benchmark to separate properties that are driven by
the statistical mechanics of the market institution from those that
are driven by conditional intelligent behavior.


\addcontentsline{toc}{section}{References and acknowledgments}
 
\begin{thebibliography}{99}

\bibitem{Bachelier00} L. Bachelier, ``Th\'{e}orie de la
  sp\'{e}culation'', 1900.  Reprinted in P.H. Cootner, \emph{The
    Random Character of Stock Prices}, 1964, MIT Press Cambridge.

\bibitem{Becker62} Becker, G., Irrational behavior and economic
  theory, \emph{J. of Political Economy} \textbf{70}, 1 (1962) 1-13.

\bibitem{Gode93} D. Gode and S. Sunder, Allocative efficiency of
  markets with zero intelligence traders: Markets as a partial
  substitute for individual rationality. \emph{J. of Political
    Economy}, \textbf{101}, 119 (1993).

\bibitem{Daniels03} Daniels, M, Farmer, J.D., Gillemot, L., Iori, G.,
  and Smith, D.E. Quantitative model of price diffusion and market
  friction based on trading as a mechanistic random
  process. \emph{Physical Review Letters} \textbf{9019}, 10 2003.

\bibitem{Smith03} Smith, E., Farmer, J.D., Gillemot, L., and
  Krishnamurthy, S., Statistical theory of the continuous double
  auction.  To appear in \emph{Quantitative Finance}, 2003.

\bibitem{Mendelson82} H. Mendelson, Market behavior in a clearing
  house, \emph{Econometrica} \textbf{50}, 1505-1524 (1982).

\bibitem{Cohen85} K.J. Cohen, R.M. Conroy and S.F. Maier, Order flow
  and the quality of the market, in: Y. Amihud, T. Ho and R. Schwartz,
  eds., \emph{Market Making and the Changing Structure of the
    Securities Industry}, 1985, Lexington Books, Lexington MA.

\bibitem{Domowitz94} I. Domowitz and Jianxin Wang, Auctions as
  algorithms, \emph{J. of Econ. Dynamics and Control} \textbf{18}, 29
  (1994).

\bibitem{Bollerslev97} T. Bollerslev, I. Domowitz, and J. Wang, Order
  flow and the bid-ask spread: An empirical probability model of
  screen-based trading, \emph{J. of Econ. Dynamics and Control}
  \textbf{21}, 1471 (1997).

\bibitem{Bak97} P. Bak, M. Paczuski, and M. Shubik, Price variations
  in a stock market with many agents, \emph{Physica A} \textbf{246},
  430 (1997).

\bibitem{Eliezer98} D. Eliezer and I.I. Kogan, Scaling laws for the
  market microstructure of the interdealer broker markets,
  http://xxx.lanl.gov/cond-mat/9808240.

\bibitem{Maslov00} S. Maslov, Simple model of a limit order-driven
  market, \emph{Physica A} \textbf{278}, 571(2000).

\bibitem{Slanina01} F. Slanina, Mean-field approximation for a limit
  order driven market model, \emph{Phys. Rev. E}, \textbf{64}, 056136
  (2001).

\bibitem{Challet01} D. Challet and R. Stinchcombe, Analyzing and
  Modeling 1+1d markets, \emph{Physica A} \textbf{300}, 285, (2001).

\bibitem{Bouchaud02} J.-P. Bouchaud, M. Mezard, M. Potters,
  Statistical properties of the stock order books: empirical results
  and models, \emph{Quantitative Finance} \textbf{2} (2002) 251-256.

\bibitem{Bouchaud03} Potters, M. \& Bouchaud, J.-P., "More statistical
  properties of orders books and price impact", \emph{Physica A} {\bf
    324}, 133-140 (2003)

\bibitem{Lillo03} Lillo,F. Farmer. J.D. \& Mantegna, R.N., Master
  Curve for Price-Impact Function, \emph{Nature} \textbf{421}, 129-130
  (2003).

\bibitem{Gabaix03} Gabaix, X., Gopikrishnan, P., Plerou, V. and
  Stanley, H.E. A theory of power-law distributions in financial
  market fluctuations, \emph{Nature} \textbf{423}, 267-270 (2003).

\bibitem{LSEbulletin} \emph{SETS four years on - October 2001},
  published by the London Stock Exchange

\bibitem{Zovko02} Zovko, I. and Farmer, J.D., The power of patience: A
  behavioral regularity in limit order placement, \emph{Quantitative
    Finance}, October 2002.

\bibitem{Bouchaud03.2} Bouchaud,J.-P., Gefen,Y., Potters,M., and
  Wyart,M.  (2003).  Fluctuations and response in financial markets:
  the subtle nature of 'random' price changes,
  xxx.lanl.gov/cond-mat/0307332.

\bibitem{Lillo03.2} Lillo, F. and Farmer, J.D., The long memory of the 
efficient market, xxx.lanl.gov/cond-mat/0311053 (2003).

\bibitem{beran} Beran, J. \emph{Statistics for Long-Memory Processes},
  Chapman \& Hall (1994).

\bibitem{Hausman92} J.A. Hausman and A.W. Lo, ``An ordered probit
analysis of transaction stock prices'', \emph{Journal of Financial
Economics} {\bf 31} 319-379 (1992)

\bibitem{Farmer96} J.D. Farmer, ``Slippage 1996'', Prediction Company
internal technical report (1996), http://www.predict.com/jdf/slippage.pdf

\bibitem{Torre97} N. Torre, \emph{BARRA Market Impact Model Handbook},
BARRA Inc, Berkeley CA, www.barra.com (1997).

\bibitem{Kempf98} A. Kempf and O. Korn, ``Market depth and order
size'', University of Mannheim technical report (1998).

\bibitem{Plerou01} V. Plerou, P. Gopikrishnan, X. Gabaix, and
H.E. Stanley, ``Quantifying stock price response to demand
fluctuations'', {Phys. Rev. E} {\bf 66} 027104 (2002).

\bibitem{Farmer03} Farmer, J.D. and Lillo, F., On the origin of power
  law tails in price fluctuations, xxx.lanl.gov/cond-mat/0309416.

\bibitem{Farmer03c} Nelkin, I., Innovations in
  trading strategies, \emph{Quantitative Finance} {\bf 3}, Number 4,
  C63-74 (2003).  (See relevant remarks by J.D. Farmer).

\bibitem{greene} William H. Greene, Econometric Analysis, Prentice Hall.

\end{thebibliography}
\end{document}